\RequirePackage{fix-cm}
\documentclass[smallextended]{svjour3}       % onecolumn (second format)
\smartqed  % flush right qed marks, e.g. at end of proof
\usepackage{graphicx}
\journalname{Foundations of Physics}

\begin{document}

\title{Are quantum spins but small perturbations of \\ ontological Ising spins?}

\titlerunning{Are quantum spins but small perturbations of ontological Ising spins?}        

\author{Hans-Thomas Elze}

\institute{Hans-Thomas Elze \at 
Dipartimento di Fisica ``Enrico Fermi'' \\ Universit\`a di Pisa, Largo Pontecorvo 3, I-56127 Pisa, Italia \\  
\email{elze@df.unipi.it}           
}

\date{Received: 11 July 2020 / Accepted: 4 August 2020}

\maketitle

\begin{abstract}
The dynamics-from-permutations of classical Ising spins is generalized here 
for an arbitrarily long chain. This serves as an {\it ontological model} with 
discrete dynamics generated by pairwise exchange interactions defining the unitary 
update operator. The model incorporates a finite signal velocity and resembles in many 
aspects a discrete free field theory. We deduce the corresponding Hamiltonian operator and show 
that it generates an exact terminating Baker-Campbell-Hausdorff formula. Motivation for this 
study is provided by the {\it Cellular Automaton Interpretation of Quantum Mechanics}. 
We find that our ontological model, which is classical {\it and} deterministic, 
appears {\it as if} of quantum mechanical kind in an appropriate formal description. 
However, it is striking that (in principle arbitrarily) small deformations of the model 
turn it into a genuine quantum theory. This supports the view that quantum mechanics stems 
from an epistemic approach handling physical phenomena.   

\end{abstract}

\section{Introduction}

New interest in a special class discrete deterministic systems has been caused by 't\,Hooft's {\it Cellular Automaton Interpretation of Quantum Mechanics} (CAI) \cite{tHooft2014,tHooft2020}. The particular systems under study are characterized by evolution of the underlying {\it ontological states} ($\cal OS$) through permutations among themselves. 

We have earlier studied simple three- and four-body models consisting of interacting two-state Ising spins, in order to illustrate the approach of CAI in some detail \cite{ElzeQu19,ElzePAFT19}. Presently, our aim is to generalize this for the case of arbitrarily long Ising spin chains.  

A basic tenet of this interpretation of quantum mechanics is that there are no superposition states ``out there''. While superpositions are an unmistakable sign of and figure prominently in the quantum theoretical description of physical reality. According to CAI, they are constructs of the mathematical language used to describe $\cal OS$ and their observed behaviour, see also Ref.\,\cite{Rovelli2015} for a lucid discussion of related issues. 

This has important implications for the foundations of quantum theory and has been amply discussed -- see Refs.\,\cite{tHooft2014,tHooft2020} for this, including numerous references to explorations of other ``pre-quantum'' models or further references, for example, in       
Ref.\,\cite{ElzeQu19}. Most importantly, widespread {\it quantum puzzles} or notions of {\it quantum weirdness} find rational explanation here in an elegant way which keeps {\it Occam's Razor} at bay. -- For the reader who is not yet familiar with this 
new interpretation of quantum mechanics, we present a {\it Synopsis of CAI} concluding this Introduction. 
 
Presently we analyse the {\it dynamics-from-permutations} of a large composite system of two-state Ising spins. This is encoded in a unitary permutation matrix, which represents pairwise exchange interactions among the spins. 

In Section\,2, some generic properties of such permutation matrices are described and related to so-called {\it cogwheel models} \cite{tHooft2014}. Making use of this information, one can obtain  Hamiltonian operators corresponding to such unitary update operators. -- We remark that all models based on a finite total number of states that evolve by permutations can be mapped on (suitable combinations of) cogwheel models \cite{tHooft2020}. Nevertheless, the physical interpretation rests on the specific states and their interactions and may involve conservation laws.  

These results are employed to derive the Hamiltonian for an arbitrarily long chain of Ising spins in Section\,3. This is a classical system evolving deterministically, but now described in a way that closely resembles quantum theory. We recall some essential properties of exchange interactions, before embarking on this derivation. 
Indeed, the extracted Hamiltonian here becomes a high-order polynomial of exchange operators acting on multi-qubit states. We comment on these aspects, also showing that the Hamiltonian generates a terminating {\it Baker-Campbell-Hausdorff formula} incorporating exchange operators. 

A first physical interpretation is offered, which brings the description close to a noninteracting classical field theory. 

In Section\,4, we argue that even (in principle arbitrarily) small perturbations of the Hamiltonian, will tend to move the representation of the dynamical system at hand from being intrinsically classical, despite quantum mechanical appearance, to become the one of a genuine quantum system. 
This includes the formation of {\it superpositions}, which can no longer be counted among the $\cal OS$ of the classical spin chain. Furthermore, we give an example showing that even 
{\it entangled states} are within reach of an approximate description of an otherwise 
ontological model. 

Present findings clearly support the view of CAI that quantum mechanics stems from an epistemic approach handling physical phenomena. This poses interesting questions and asks for model building aimed at more realistic theories.  

The relevance of Ising models for unravelling the content of Bell's theorem and statistical issues in quantum theory, other than serving as our model for evolving $\cal OS$, has recently also been discussed, for example, in Refs.\,\cite{Vervoort1,Vervoort2,Wetterich16a,Wetterich16b}. 

Section\,5 provides our conclusions.  

\subsection{A synopsis of the Cellular Automaton Interpretation} 

The new interpretation of quantum mechanics (CAI) is based on distinguishing {\it three kinds of 
states}, which we introduce in turn. 

Quantum states here form part of the mathematical language used and, thus, are of epistemological character \cite{Rovelli2015}. They are ``templates'' for the description of the reality of ontological states beneath, which evolve deterministically \cite{tHooft2014,tHooft2020}.   

\vskip 0.1cm \noindent 
{\it \underline{Definition}.} \hskip 0.1cm
Ontological States ($\cal OS$) are states a closed physical system can be in. 
\vskip 0.1cm

\noindent 
The set of $\cal OS$ may be very large, possibly even infinite. For now, it is assumed to be denumerable. 

\vskip 0.1cm \noindent 
{\it \underline{Hypothesis}.} \hskip 0.1cm
Superpositions of $\cal OS$ do not exist in physical reality ``out there''.  
\vskip 0.1cm

\noindent 
This hypothesis seems plainly wrong, in view of the overwhelming role played by superpositions in quantum theory. However, we anticipate and stress that quantum superpositions ``happen'' in the realm of the language used, {\it i.e.}, the mathematical formalism describing observed phenomena. 

\vskip 0.1cm \noindent 
{\it \underline{Corollary}.} \hskip 0.1cm
The $\cal OS$ evolve by {\it permutations} among themselves.\footnote{Denoting $\cal OS$ by  $|A\rangle ,\; |B\rangle ,\; |C\rangle ,\; |D\rangle ,\;\dots\;$, for example, such a dynamics could be simply this:
\begin{equation} \label{perms}   
|A\rangle\rightarrow |D\rangle\rightarrow |B\rangle\rightarrow\;\dots\;\;.
\end{equation} 
This kind of evolution is the only possible one, unless the set of states itself changes, {\it i.e.}, grows or shrinks. We do not consider a changing set of states here, however, this could be interesting when it comes to the evolving Universe.}   
\vskip 0.1cm

\vskip 0.1cm \noindent 
{\it \underline{Definition}.} \hskip 0.1cm
The $\cal OS$ are declared to form a fixed orthonormal set (fixed once for all) and a {\it Hilbert space} ${\cal H}$ is defined with respect to this preferred basis.\footnote{Diagonal operators on this basis are {\it beables} and their eigenvalues characterize physical properties of the states, corresponding to the labels $A,\; B,\; C,\;\dots\;$ used in the previous footnote.} 
\vskip 0.1cm

\noindent 
Association of the special Hilbert space ${\cal H}$ with the set of $\cal OS$, leads us to the following definition. 

\vskip 0.1cm \noindent 
{\it \underline{Definition}.} \hskip 0.1cm 
Quantum States ($\cal QS$) 
are formal superpositions of $\cal OS$ defined in ${\cal H}$. 
\vskip 0.1cm

\noindent
These are the ``templates'' for doing  physics with the help of mathematics.\footnote{The amplitudes that specify a $\cal QS$ need interpretation, when describing experiments. By experience, relating amplitudes to probabilities has been extraordinarily useful. In this way, the {\it Born rule} is built in by definition! -- The Born rule can also be understood as a counting procedure related to a conserved two-time function of Hamiltonian cellular automata, which generalizes the norm of $\cal QS$ \cite{FFM17}. -- It is is not forbidden by any element of quantum theory to abandon the proportionality between {\it absolute values squared} of complex amplitudes and probabilities, however, at the price of unnecessarily complicating its mathematical tools \cite{tHooft2014}.}      

It will often be difficult to relate unitary evolution of $\cal OS$ by permutations to a familiar looking Hamiltonian operator, in particular in presence of interactions \cite{tHooft2014,tHooft2020,ElzeQu19,ElzePAFT19,Wetterich16a,Wetterich16b,FFM17}. Which makes the accessible spin chain model of Section\,3 interesting. In any case, we see the machinery of quantum theory already largely in place, including important techniques of unitary transformations in Hilbert spaces, but with the new perspective offered through $\cal OS$ existing ``out there''. 

Also {\it classical states} need to be defined in relation to $\cal OS$.\footnote{Usually, they are thought to describe limiting situations of quantum mechanics, {\it e.g.} in presence of environment induced decoherence. They constitute the realm of classical physics.} For CAI, classical states belong to deterministic macroscopic systems, including billiard balls, pointers of apparatus, planets, etc., {\it i.e.}, situations where large numbers of ontological states must be involved. 

\vskip 0.1cm \noindent 
{\it \underline{Definition}.} \hskip 0.1cm 
Classical States ($\cal CS$) 
%\\ \phantom .\hskip 2.6cm 
of physical (macro-)systems are the  
%\\ \phantom .\hskip 2.6cm 
probabilistic distributions of $\cal OS$ which are not resolved individually. 
\vskip 0.1cm 

Repeatedly performed experiments or approximately repeating evolutions of a sufficiently (but never completely) isolated part of the Universe   
pick up different initial conditions regarding $\cal OS$. 
Therefore, a {\it classical apparatus} forming part of such a situation must be expected to yield different pointer positions as outcomes. Furthermore, the  
probability of a particular outcome directly reflects the probability of having a particular $\cal OS$ as initial condition. 

Thus, as a consequence of the absence of superpositions of $\cal OS$ and of their evolution by permutations among themselves, we obtain a conservation law:\,\footnote{Using quantum superpositions of $\cal OS$ to describe the initial state approximately, we obtain for an evolving $\cal QS$ 
$|Q\rangle$: 
\begin{eqnarray} \label{QS} 
|Q\rangle &:=&\alpha |A\rangle +\delta |D\rangle +\dots  
\;,\;\;\;|\alpha |^2+|\delta |^2 +\dots\; =1
\;\;, 
\\ [1ex] \label{QSEvol}
\mbox{then},&\phantom .& 
|Q\rangle\;\longrightarrow\;
\alpha |D\rangle +\delta |B\rangle +\dots  
\;\;. \end{eqnarray} 
We see that the {\it amplitudes remain the initial ones}, while the $\cal OS$ evolve by 
permutations, in the chosen example according to (\ref{perms}) of footnote\,$^1$.} 

\vskip 0.1cm \noindent 
{\it \underline{Corollary}.} \hskip 0.1cm
The {\it Conservation of Ontology}. 
\vskip 0.1cm

\noindent 
Hence, the reduction or collapse to a $\delta$-peaked distribution of pointer positions, the core of the {\it measurement problem}, is an apparent effect. It arises due to the intermediary use of {\it quantum mechanical templates}, in particular superposition states, when describing the evolution of what in reality are $\cal OS$ that differ in different runs of an experiment. 

According to CAI, superpositions of $\cal OS$ do not exist 
``out there'' \cite{tHooft2014}. There is nothing to which an $\cal OS$ could be reduced and, 
therefore, {\it no collapse} can occur!\,\footnote{This does not imply that quantum mechanical superposition states are to be avoided. On the contrary, part of the motivation for CAI is to better understand, why they are so extremely effective in describing experiments probing nature.} 
  
It is encouraging for this ontological theory that no stochastic or nonlinear reduction process 
is needed, which would modify its collapse-free linear unitary evolution. Measurements, according to CAI, are {\it interactions} between the degrees of freedom belonging to an object and those belonging to an apparatus, altogether evolving through ontological states. 

The elegance and simplicity of this view is very appealing. However, it is not entirely straightforward to construct examples of interacting systems and only quite recently a general scheme has been proposed for this purpose \cite{tHooft2020}.  

We conclude here by underlining the emphasis that CAI puts on the distinction between  
{\it ontic and epistemic aspects} (of states) encountered in the complex situations, where quantum theory is usually invoked. This should be quite relevant also, for example, when 
the formalism of quantum theory is successfully applied to situations considered outside of physics \cite{AtmanspacherPrimas,Khrennikov2}. It may even be fruitful sometimes to view ontic and epistemic features side by side, such as represented for example by the conformational and functional aspects of complex molecules \cite{Khrennikov1}. Last not least, the longstanding question whether classical (ontological) and quantum mechanical degrees of freedom can coexist  consistently in one dynamical framework deserves reconsideration in the light of CAI, see Ref.~\cite{Elzehybrid} and earlier references therein.\footnote{On a historical note, we add here that the discussion of ontological {\it vs.} epistemological approaches to the building of theories of physics has an intense precursor in times when Newtonian physics was superseded by field theories of forces in the sequel of Maxwell's electrodynamics \cite{Hertz,Boltzmann}. The fine distinction between qualitatively different types of theories or theory building seems to have been lost during the rapid developments leading to Quantum Mechanics in the following. 
Attention to this has been drawn by Khrennikov, indicating  possible consequences for the persisting interpretational problems of quantum theory \cite{KhrennikovOld}.}

\section{Complete permutations of $N$ objects}   

In the following we will make use of some basic properties of permutations of $N$ objects, which we collect here for completeness; see also Refs.\,\cite{ElzeQu19,ElzePAFT19}. 

\subsection{Complete permutations and their Hamiltonian operators}
The objects to be considered are the {\it states} or configurations of a classical Ising spin model that evolve by permutations among themselves. -- Of particular interest are the {\it complete permutations} that map  
the $N$ states, say $\alpha_1,\alpha_2,\;\dots \;,\alpha_N$, in precisely $N$ steps onto one another, involving {\it all states once}. 

Complete permutations are represented by {\it unitary} $N\times N$ matrices,  
$\hat U_N$, which have exactly {\it one} off-diagonal arbitrary phase 
per column and row, $\exp (i\phi_k), k=1,\;\dots \;,N$, and vanishing matrix elements elsewhere. 

Such a permutation matrix can be given in a {\it standard form}: 
\begin{equation} \label{UN} 
\hat U_N:= 
\left (
\begin{array}{c c c c c c} 
0           & .           & .           & .     & 0       & e^{i\phi_N} \\ e^{i\phi_1} &0            & .           &       & .       & 0  \\ 
0           & e^{i\phi_2} & 0           & .     &         & .  \\
.           & 0           & e^{i\phi_3} & 0     &         & .  \\ 
.           &             & .           & .     & .       & .  \\
0           & .           & .           & 0     & \;e^{i\phi_{N-1}} & 0  \\
\end{array}\right )
\;\;,\;\;\; 
\hat U_N\hat U_N^\dagger =\mathbf{1} 
\;\;,
\end{equation}
corresponding to an appropriate ordering of the $N$ states (cf. the {\it auxiliary basis} 
introduced in Section\,2.2).  

It is easy to see that for all complete permutation matrices holds: 
\begin{equation} \label{UNN} 
(\hat U_N)^N=e^{i\sum_{k=1}^N \phi_k}\;\mathbf{1}
\;\;. \end{equation} 
This implies that their eigenvalues lie on a unit circle in the complex plane and are given 
by the $N$th roots of 1, multiplied by an overall phase. 	

Next, we may define a related {\it Hamiltonian} by: 
\begin{equation} \label{Hop} 
\hat U_N=:e^{-i\hat H_NT} 
\;\;, \end{equation} 
with $T$ a fixed time scale. 
The eigenvalues of $\hat H_N$ are then obtained directly from Eq.\,(\ref{UNN}), 
which yields the diagonalized Hamiltonian:   
\begin{equation} \label{Hdiag} 
\hat H_{N,\;diag}=\mbox{diag}\Big (\frac{1}{NT}\big (2\pi (n-1)
-\sum_{k=1}^N\phi_k\big )\;|\;n=1,\;\dots\; ,N
\Big ) 
\;\;. \end{equation} 
The phases are irrelevant for us at present, 
hence we set $\phi_k\equiv 0$, from now on. 

So far, we determined the eigenvalues of complete permutation matrices and, thus, 
obtained the diagonal form of the Hamiltonian of Eq.\,(\ref{Hdiag}). As before in 
Refs.\,\cite{ElzeQu19,ElzePAFT19}, where small systems composed of three and four Ising 
spins were considered, we are interested in  
the Hamiltonian for the standard form of $\hat U_N$, Eq.\,(\ref{UN}), {\it i.e.},  
in the case of $N$ states.   

\subsection{Eigenvectors, diagonalizing matrix, and Hamiltonian for complete permutations in 
standard form}  

The standard form of a complete permutation matrix $\hat U_N$, Eq.\,(\ref{UN}), refers to the  
basis of normalized {\it auxiliary vectors} defined by: 
\begin{equation} \label{AuxVec} 
|\alpha_m\rangle :=(0,\dots ,0,1,0,\dots ,0)^t\;\;,\;
m=1,\dots ,N
\;\;, \end{equation} 
with entry 1 at the $m$-th position, respectively. Then,   
we make the following {\it Ansatz} for the normalized eigenvectors, $|A_n\rangle ,\; n=1,\;\dots \;,N$: 
\begin{equation} \label{eigenvec} 
|A_n\rangle =\frac{1}{\sqrt N}\sum_{m=1}^N
e^{ia_{nm}}|\alpha_m\rangle 
\;\;, \end{equation} 
which incorporates the auxiliary basis. This   
amounts to a discrete Fourier transformation and is suggested by explicit examples for small $N$. 
Setting the timescale to $T=1$, for a moment, and inserting the 
Ansatz into $\hat U_N|A_n\rangle =\exp (-iE_n) |A_n\rangle$, with $E_n=2\pi (n-1)/N$, gives a recursion relation for the phases $a_{nm}$ \cite{ElzeQu19}. 
%: 
%\begin{equation} \label{recursion} 
%a_{nm+1}=a_{nm}+E_n\;\;(\mbox{mod}\;2\pi)\;\;,\;\; 
%a_{n1}=a_{nN}+E_n\;\;,\;\;a_{n1}:=0 
%\;\;. \end{equation}  
Solving this, we obtain: 
\begin{eqnarray} \label{phases} 
a_{nm}&=&\frac{2\pi}{N}(nm-n-m+1)\;\;(\mbox{mod}\;2\pi)  
\\ [1ex] \label{phasessymm}
&=&a_{mn}
\;\;, \end{eqnarray} 
with $a_{n1}:=0$; and, thus, we find the eigenvectors. 

These phases also fix the unitary 
{\it diagonalizing matrix} $\hat D$ which maps auxiliary basis vectors 
to eigenvectors, as can be read off from our Ansatz (\ref{eigenvec}): 
\begin{equation} \label{diagonalM} 
|A_n\rangle =\sum_{m=1}^N(\hat D)_{nm}|\alpha_m\rangle\;\;,\;\;
(\hat D)_{nm}:=\frac{1}{\sqrt N}e^{ia_{nm}}
\;\;. \end{equation} 
  
The diagonalizing matrix $\hat D$ is needed, in order to relate the diagonalized Hamiltonian, $\hat H_{N,\;diag}$ of Eq.\,(\ref{Hdiag}), to its generic form defined through Eq.\,(\ref{Hop}): 
\begin{equation} \label{Hgeneric} 
\hat H_N=\hat D^\dagger \hat H_{N,\; diag}\hat D 
\;\;. \end{equation} 
Calculating the matrix elements of $\hat H_N$ from this relation yields: 
\begin{eqnarray} \label{HgenericM1} 
(\hat H_N)_{nn}&=&\frac{\pi}{NT}(N-1)\;\;,\; n=1,\;\dots\; ,N 
\;\;, \\ [1ex] \label{HgenericM2} 
(\hat H_N)_{n\neq m}&=&\frac{\pi}{NT}
\Big (-1+i\cot\big (\frac{\pi}{N}(n-m)\big ) \Big ) 
\;\;,\; n,m=1,\;\dots\; ,N  
\;\;. \end{eqnarray} 
Thus, we have $\hat H_N=\hat H_N^\dagger$, as it should be. 

We remark that the matrix elements of $\hat H_N$ are {\it constant} along lines parallel to the diagonal and on the diagonal. Therefore, matrix 
elements of adjacent rows or columns differ by {\it cyclic permutation} of the entries only. -- 
We shall make use of this important property shortly.   

\subsection{Permutations and cogwheel models} 

The results represented in the preceding paragraphs underlie so-called 
{\it cogwheel models} \cite{tHooft2014}, where a single degree of freedom moves in discrete 
steps periodically through a finite number of states, {\it e.g.} equidistant positions of a particle on a circle. These are   
deterministic ``classical'' models with time reversible quantum mechanical features. 
In suitable limits, namely $N\rightarrow\infty$ and $T\rightarrow 0$, 
with $NT\equiv\omega^{-1}$ fixed, the  
{\it quantum harmonic oscillator} is described by such a model, while for $\omega\rightarrow 0$ the {\it free quantum particle} results \cite{tHooft2014,tHooft2020,Elze,ElzeRelativPart}. 

At this point, the analysis of complete permutations of the standard form given in Eq.\,(\ref{UN}), or of {\it cogwheel models}, seems complete. This may give the impression that only a very limited kind of ontological models of physical systems can be constructed along these lines, where quantum mechanical features emerge from deterministic classical dynamics. 

In particular, it has been unclear for a long time, whether it is possible 
at all to introduce interactions among the simplest building blocks, 
which are ontological models related to quantum mechanical one-body systems \cite{PRA2014,Wigner13}, such as free field modes or free particles. 

Yet a recent proposal by 't\,Hooft addresses this issue, trying to couple different cogwheels \cite{tHooft2020}. --  
In another attempt to have access to more complex situations, we instead have studied  
small composites of Ising spins which are coupled by 
exchange interactions \cite{ElzeQu19,ElzePAFT19}. We will generalize 
this here for an arbitrarily large number of such classical spins in a chain. 
Higher-dimensional arrays are treated similarly. 

\section{Dynamics-from-permutations from Ising spin exchange} 

The hypothesis of {\it ontological states} ($\cal OS$) existing ``out there'', which evolve  deterministically, interprets why and how quantum theory describes successfully the reality 
abstracted from experiments or observations. Which is the essence of the 
{\it Cellular Automaton Interpretation of Quantum Mechanics} \cite{tHooft2014}. 

We intend to further explore now the potential of {\it dynamics-from-permutations}, as already indicated in Section\,2, to lead to complex phenomena and some of its interesting formal aspects. 

Keeping the established cogwheel models in mind, we need to further examine how permutations are acting there and expressed in relation to auxiliary basis vectors, {\it cf.} Eqs.\,(\ref{UN}) and (\ref{AuxVec}). Note that the dimension of the $N\times N$-matrix $\hat U_N$ can be varied, but our formulation applies equally well for small systems as for systems with $N>>1$. 

Nothing has been said in more physical terms, so far, about {\it what constitutes $\cal OS$} and about 
{\it why or how they evolve by permutations} among themselves. In the following, this will be pursued one step further in the context of classical {\it Ising spin} or equivalent {\it bit processing} models.   

\subsection{Exchange operations, transpositions and permutations} 
 
Consider a one-dimensional chain consisting of $2S+1,\; S\in\mathbf{N}$, classical {\it two-state Ising spins}, labeled ``$1,2,3,\dots ,2S+1$''. For simplicity, we assume {\it periodic boundary conditions}, which are implemented by identifying the last with the first spin.\footnote{The following considerations work with other boundary conditions as well. However, periodic boundary conditions turn out to yield the most transparent picture of the dynamics.} This is equivalent to having $2S$ Boolean variables or {\it bits}. They can be in one of $2^{2S}$ states. -- 
We emphasize that such multi-spin states as $\cal OS$, by definition, physically do {\it not form superposition states}. Mathematically, of course, we are free to do that ``by 
mistake'', as will be discussed in Section\,4.   

We are interested to couple the $2S$ spins in a simple way that leads to permutations among their $2^{2S}$ states. They can indeed be generated by {\it spin exchange}, a permutation which is a {\it transposition} of two spins, $\hat P_{ij}\;(\equiv\hat P_{ji})$, $i,j=1,\;\dots\;,2S$, with these properties: 
\begin{equation} \label{Pijdef} 
\hat P_{ij}|s_i,s_j\rangle :=
|s_j,s_i\rangle 
\;\;,\;\;\; \hat P_{ji}\hat P_{ij}
=(\hat P_{ij})^2=\mathbf{1}
\;\;, \end{equation} 
where the states of a single spin can assume the values    
$s_k=\pm 1$ or, equivalently, $s_k=\uparrow ,\downarrow$, {\it i.e.} ``spin up, spin down'', respectively; we use the ket notation $|s_i,s_j\rangle$ 
to indicate that the first spin has value $s_i$, the second value $s_j$, and continuing in this way for a many-spin state. 

To familiarize ourselves with this setting, it may be useful to consider additional symmetry properties or {\it conservation laws} 
concerning permutations. 

An obvious property of {\it all} permutations here is the {\it separate conservation of numbers of up and down spins}:  
\begin{equation} \label{Nupdown} 
[\hat N_u,\hat P_{ij}]=[\hat N_d,\hat P_{ij}]=0\;\;,\;\mbox{for all}\; i,j 
\;\;, \end{equation} 
where the respective number operators can be defined explicitly in terms of the Pauli matrix $\hat\sigma^z$. 
In fact, identifying the two states $s_k=\pm 1$ of Ising spin ``$k$'' with the eigenstates of the 
Pauli matrix $\hat\sigma_k^{\; z}$, $\psi_+=(1,0)^t$ and $\psi_-=(0,1)^t$, respectively, we find that   
the above number operators are $\hat N_u:=S+\sum_{k=1}^{2S}\hat\sigma_k^{\; z}/2$ and 
$\hat N_d:=S-\sum_{k=1}^{2S}\hat\sigma_k^{\; z}/2$.

This implies, that the {\it total number} of spins is conserved: 
\begin{equation}\label{Ntotal} 
\hat N:=\hat N_u+\hat N_d=2S   
\;\;, \end{equation}  
of course. Furthermore, defining the {\it magnetization} of the chain by:  
\begin{equation}\label{magnet}
\hat M:=(\hat N_u-\hat N_d)/(\hat N_u+\hat N_d)=\sum_{k=1}^{2S}\hat\sigma_k^{\; z}/2S 
\;\;, \end{equation} 
this is conserved as well.  
 
These properties have the important consequence that not all of the $2^{2S}$ states can be reached from a given initial state, no matter how we define the dynamics in terms of exchange operators. 

Furthermore, for any state with $N_u$ up- and $N_d$ down-spins and for all sequences of permutations acting on it, there is a state with the numbers of up- and down-spins exchanged, 
$N_u'=N_d$ and $N_d'=N_u$, which evolves under the same permutations in one-to-one correspondence with the former one. This means, the total {\it spinflip  operator}, symbolically $\hat C:\;\uparrow\leftrightarrow\downarrow$, commutes with  exchange operations:
\begin{equation} \label{C} 
[\hat C,\hat P_{ij}]=0\;\;,\;\mbox{for all}\; i,j 
\;\;. \end{equation} 
Also this spin-flip operator can be defined in terms of Pauli matrices, namely: 
$\hat C:=\prod_{k=1}^{2S}\hat\sigma_k^{\; x}$. 

Finally, we recall that the unitary operator $\hat P_{ij}$ can be expressed as well in terms 
of the Pauli spin-1/2 matrices: 
\begin{equation} \label{PijPauli}
\hat P_{ij}=\frac{1}{2}(\underline{\hat\sigma}_i
\cdot\underline{\hat\sigma}_j+\mathbf{1})
\;\;, \end{equation} 
where $\underline{\hat\sigma}$ is a vector formed by the   
components $\hat\sigma^x,\hat\sigma^y,\hat\sigma^z$. 
 
These expressions indicate some relation with the QM of {\it spin-1/2} entities or {\it qubits}, mentioned before \cite{ElzeQu19}. We shall continue to discuss this in the following.

Concluding this section, we have to keep in mind that two permutations involving three different spins do not commute: 
\begin{equation} \label{comm} 
[\hat P_{ij},\hat P_{jk}]\neq 0\;\;, \;\;
\mbox{for}\;\; i\neq k
\;\;, \end{equation} 
without summation over $j$. Note that for $a,b,c=\pm 1$, arbitrary but fixed,  
one obtains, {\it e.g.}, $\hat P_{12}\hat P_{23}|abc\rangle =|cab\rangle\neq |bca\rangle =
\hat P_{23}\hat P_{12}|abc\rangle$. It is this   
{\it noncommutativity} which forces us to perform additional steps in order to extract the  Hamiltonian from a unitary update operator, defined in terms of transpositions of Ising spins among a collection of such classical variables.   

\subsection{Unitary dynamics of an Ising spin-chain model} 

We shall define a particular unitary operator $\hat U$ that evolves the states of $2S$ classical two-state spins in a discrete time step $T$. Here we make use of the exchange operations discussed in the preceding section. We will then proceed to extract the corresponding Hamiltonian $\hat H$, employing results for complete permutations that we obtained before, see Eqs.\,(\ref{Hop})--(\ref{Hdiag}) or (\ref{Hgeneric}).  

The dynamics of our spin chain model is defined by the following product of exchange operations: 
\begin{equation} \label{U} 
\hat U:=\prod_{k=1}^S\hat P_{2k-1\;2k}\prod_{l=1}^{S}\hat P_{2l\;2l+1}
=:\exp (-i\hat HT) 
\;\;, \end{equation} 
acting sequentially from right to left on the pairs of spins indicated by the respective indices on the exchange operators; because of the periodic boundary condition we have  
to identify $\hat P_{2S\;2S+1}\equiv\hat P_{2S\;1}$. 
Furthermore, notice that we have separated the pairs of spins that occur into two groups. Which we call {\it even} and {\it odd pairs}, respectively, depending on whether 
the first index on a given $\hat P_{ij}$ is even or odd; despite the fact that 
$\hat P_{ij}=\hat P_{ji}$, formally, we now keep the indices ordered according to $\hat P_{i<j}$, 
to be definite. 

The splitting of the update operator $\hat U$ into a product of (at least) two groups 
of operators that {\it commute} among themselves, as above, implements a {\it finite signal velocity} in the model. Otherwise, if {\it all} 
pairs of spins would be updated sequentially, a local change of the state, say flipping one 
Ising spin, would generally be felt within one time step $T$ everywhere in an 
arbitrarily long chain.   

The finite signal velocity becomes visible, if we follow the update of an arbitrary but fixed state through several steps. With the ket notation as before, we denote this {\it initial state} by: 
\begin{equation}\label{psi} 
|\psi\rangle :=|s_1,s_2,s_3,\dots ,s_{2S-2},s_{2S-1},s_{2S}\rangle 
\;\;, \end{equation} 
where each variable has a given value, either $s_k=1$ or $s_k=-1$, {\it i.e.}, spin ``1'' has value $s_1$, spin ``2'' has value $s_2$, {\it etc.} In this way numbering the spins  
or initial positions of the variables within the ket consecutively from left to right, 
we call them {\it leftmoving} and {\it rightmoving positions}, respectively, according their number being odd or even. These names are suggested by observing the simple dynamics resulting from consecutive updates under the periodic boundary condition: 
\begin{eqnarray}\label{U1psi} 
\hat U^1|\psi\rangle &=&|s_3,s_{2S},s_5,s_2,\dots ,s_{2S-1},s_{2S-4},s_1,s_{2S-2}\rangle 
\;\;, \\ [1ex] \label{U2psi} 
\hat U^2|\psi\rangle &=&|s_5,s_{2S-2},s_7,s_{2S},\dots ,s_1,s_{2S-6},s_3,s_{2S-4}\rangle 
\;\;, \\ [1ex] \label{U3psi}     
\hat U^3|\psi\rangle &=&|s_7,s_{2S-4},s_9,s_{2S-2},\dots ,s_3,s_{2S-8},s_5,s_{2S-6}\rangle
\;\;, \\ \nonumber 
&\dots & 
\\ \label{Uspsi}
\hat U^S|\psi\rangle &=&|\psi \rangle 
\;\;, \end{eqnarray} 
such that after three updates, for example, spin ``1'' has value $s_7$, spin ``2'' has value 
$s_{2S-4}$, $\dots$, while spin ``2S'' now has been updated to the value $s_{2S-6}$ from the initial state. 

We find in this way that the initial 
values of the variables, $s_1,\dots ,s_{2S}$, jump {\it two} spin positions per update either {\it leftmoving} or {\it rightmoving} along the chain, depending on their odd or even origin in the initial configuration, which is reflected in their odd or even indices. Thus, the {\it direction and velocity of motion are conserved}. Therefore, having $2S$ spins with periodic boundary condition, it takes $S$ updates to recover the initial state, as in 
Eqs.\,(\ref{psi})--(\ref{Uspsi}). 

Since the system moves periodically through (at most) $S$ states, which ones precisely being determined by the initial condition, we can make use of the results of Section\,2, in particular of Section\,2.2, in order to extract a corresponding Hamiltonian. We will follow here a similar strategy as for the 3- and 4-spin chains considered in Refs.\,\cite{ElzeQu19,ElzePAFT19}. 

The Eqs.\,(\ref{psi})--(\ref{Uspsi}) show that an arbitrary initial state 
$|\psi\rangle$ of the chain evolves according to the pattern of a cogwheel model with 
generally $S$ states, 
{\it cf.} Section\,2.3. Viewing it abstractly from internal structure and dynamics of the spin states, such evolution  
is determined by the Hamiltonian $\hat H_{N=S}$ of Eqs.\,(\ref{HgenericM1})--(\ref{HgenericM2}). 
The unitary operator $\hat U_S=\exp (-i\hat H_ST)$ maps states of the auxiliary basis, see Eq.\,(\ref{AuxVec}), sequentially onto each other in the same way as $\hat U$ does in Eqs.\,(\ref{psi})--(\ref{Uspsi}). Thus, 
we can identify evolution of the cogwheel with the one of the spin chain states:   
\begin{equation}\label{identification}
\hat U_S^{\; m-1}|\alpha_1\rangle =|\alpha_m\rangle\equiv\hat U^{m-1}|\psi\rangle
\;\;,\; m=1,\dots ,S
\;\;,  \end{equation}    
keeping in mind that $(\hat U_S)^S=\mathbf{1}$. 

In order to obtain the relevant Hamiltonian $\hat H$ 
for the spin chain, defined in Eq.\,(\ref{U}), we now have to `dress' $\hat H_S$ by appropriate (commuting) operators. Such that $\hat H$ acts on spin states, Eq.\,(\ref{psi}), in complete analogy to $\hat H_S$ acting on `bare' auxiliary basis states, Eq.\,(\ref{AuxVec}), of the cogwheel model. Generalizing what we have learnt from the cases of three or four spins \cite{ElzeQu19,ElzePAFT19} with the help of the matrix elements from Eqs.\,(\ref{HgenericM1})--(\ref{HgenericM2}), we thus obtain here: 
\begin{eqnarray}\label{H1} 
\hat H&=&\sum_{n=1}^S(\hat H_S)_{n1}\hat U^{n-1} 
\;\; \\ [1ex] \label{H2} 
&=&\frac{\pi}{T}\Big (\mathbf{1}-\frac{1}{S}\sum_{n=0}^{S-1}\hat U^{n}
+\frac{i}{S}\sum_{n=1}^{S-1}\cot (\frac{\pi}{S}n )\hat U^{n}\Big ) 
\;\; \\ [1ex] \label{H3}  
&=&\frac{\pi}{T}\Big (\mathbf{1}
+\frac{i}{S}\sum_{n=1}^{S-1}\cot (\frac{\pi}{S}n )\hat U^{n}\Big )
\;\;,\;\mbox{for}\;\;\hat U|\psi\rangle\neq |\psi\rangle 
\;\; \\ [1ex] \label{H4} 
&=&\frac{\pi}{T}\Big (\mathbf{1}
+\frac{i}{2S}\sum_{n=1}^{S-1}\cot (\frac{\pi}{S}n )
\big (\hat U^{n}-(\hat U^\dagger )^{n}\big )\Big )
\;\;. \end{eqnarray} 
We remark that the geometric series in Eq.\,(\ref{H2}) vanishes and gives Eq.\,(\ref{H3}), since 
$\hat U^S=\mathbf{1}$, provided $\hat H$ is applied to states $|\psi\rangle$ that are {\it not} 
eigenstates of $\hat U$ with {\it eigenvalue one}. 

Instead, for states with $\hat U|\psi_0\rangle =|\psi_0\rangle$, we find that $\hat H|\psi_0\rangle =0$, consistently with the fact that they do not evolve, {\it i.e.}, they are  static {\it zero modes} -- examples of these have all spins either up or down or all leftmovers up (down) and all rightmovers down (up). Note that by symmetry of the cot-function we have $\sum_{n=1}^{S-1}\cot (\frac{\pi}{S}n)=0$, in this case. 

We use this same symmetry to rewrite the final result for $\hat H$ in Eq.\,(\ref{H4}) in a 
manifestly self-adjoint form, taking into account that $\hat U^{S-k}=(\hat U^\dagger )^k$, 
which follows 
from $\hat U^k\hat U^{S-k}=\mathbf{1}=\hat U^k(\hat U^\dagger )^k$, for $1\leq k\leq S$.  

Thus, we succeed to derive the Hamiltonian $\hat H$ for the Ising spin-chain 
model, as defined in Eqs.\,(\ref{U}). Some comments are in order here, which we shall present in the following.  

\subsection{Comments on Hamiltonian, dynamics, and formal aspects of the classical spin-chain model} 

\subsubsection{Degeneracy, magnetization, and translation invariance} 

In Refs.\,\cite{ElzeQu19,ElzePAFT19} we listed all states of the short-chain models there and classified them according to conservation laws, {\it cf.} Subsection 3.1 here. This allowed to construct the Hamiltonian from its inferred block diagonal structure. Presently, instead, for a chain with a very large number of Ising spins, $2S\gg 1$, and consequently exponentially large number, {\it i.e.} $2^{2S}$, of states, we followed the evolution of an arbitrary initial  
state instead, Eqs.\,(\ref{psi})--(\ref{Uspsi}), in order to extract the Hamiltonian. Interestingly, we find that 
at most $S$ updates are needed to come back to any initial state. This implies that the Hamiltonian has 
at most $S$ different eigenvalues with, consequently, {\it exponentially growing degeneracy}. 
This may include states that evolve and come back to the initial one after less 
than $S$ steps, as seen explicitly before \cite{ElzeQu19,ElzePAFT19} -- we also mentioned the  cases of zero modes after Eqs.\,(\ref{H1})--(\ref{H4}).  

The high degeneracy of the states can be alleviated by adding a term proportional to the 
{\it magnetization} $\hat M$, Eq.\,(\ref{magnet}), to the Hamiltonian $\hat H$, with which it commutes term by term. 
However, the magnetization can only assume $2S+1$ different values. Therefore, it suffices   
to diminish only mildly the exponential in $S$ degeneracy of states.  

We have observed before that during consecutive updates, Eqs.\,(\ref{U1psi})--(\ref{Uspsi}), of a generic state, $|s_1,\dots ,s_{2S}\rangle$,  
the initial values of the variables $s_j$ move in constant steps of two units per update either to the left or to the right, for label $j$ odd or even, respectively. This corresponds to the  
{\it discrete translation invariance} of the spin-chain model: The numbering of all spins along the chain can be shifted by an even integer (mod $2S$), without affecting the 
dynamics.\footnote{Shifting by an odd integer, instead, converts leftmovers and rightmovers into each other.}  

\subsubsection{A first interpretation} 

It is tempting to associate this in every aspect discrete Ising spin system with classical degrees of freedom that describe the free motion of {\it `massless particles'} with two internal states. A left- or right-moving up-spin, for example, appears to move with constant 
velocity, {\it i.e.} the signal velocity or {\it `velocity of light'} of the model, on a lightcone 
in 1+1 dimensions, with the spacelike direction compactified by the periodic boundary 
condition. 

We remark that the {\it `center-of-mass'} of a composite, say one leftmoving and two rightmoving 
up-spins, with all other spins down, moves within such a lightcone. While 
the contribution to the magnetization or to a corresponding energy of $n$ parallel moving 
up-spins, $E(n)$ for example, is simply additive, $E(n)=nE(1)$, with $0\leq n\leq S$.  

In this picture, the underlying spin exchange interactions have been subsumed in the hopping from 
sites to either left or right next-to-nearest-neighbour sites. The left- and rightmovers do not interact and up or down spins are never flipped. 

Appropriately modifying the notation, we rewrite Eq.\,(\ref{psi}) simply as: 
\begin{equation}\label{psinew} 
|\psi\rangle\equiv |\psi_0\rangle =|\{ s_0^R(2k)\},\{ s_0^L(2l-1)\}\rangle \;\;,\; k,l=1,\dots,S 
\;\;, \end{equation} 
{\it i.e.}, collecting separately left- and rightmoving spin variables of a generic initial state. Then, 
the evolution of such a state from the $n$-th update to the $(n+1)$-st update is described {\it exactly} by the pairs of equations: 
\begin{eqnarray}\label{eq1} 
s_{n+1}^R(2k)-s_n^R(2k)&=&-\big (s_n^R(2k)-s_n^R(2k-2)\big ) 
\;\;,\; k=1,\dots,S 
\;\;, \\ [1ex]\label{eq2} 
s_{n+1}^L(2l-1)-s_n^L(2l-1)&=&s_n^L(2l+1)-s_n^L(2l-1) 
\;\;,\; l=1,\dots,S 
\;\;, \end{eqnarray} 
where we suitably subtracted identical terms on both sides, respectively. These {\it finite 
differences equations} can be mapped one-to-one on continuous space-time differential equations for bandwidth-limited classical fields $s^L(x,t)$ and $s^R(x,t)$ using {\it Sampling Theory},  
as in Ref.\,\cite{PRA2014}, for example. We shall postpone the discussion of the ensuing wave equations in the {\it continuum limit}.  

It seems possible to extent the model presented here 
to 2+1 or 3+1 dimensions along these lines. Of course, continuous rotational invariance will then be broken to a discrete subgroup.    

\subsubsection{The Baker-Campbell-Hausdorff formula behind} 

The algebraic problem of relating exponentials of noncommuting operators to each other is familiar from quantum theory or Lie group theory, with many applications, {\it e.g.}, in quantum optics or particle physics models. -- 
Consider 
$\exp (X)\exp (Y)=\exp (Z)$. The formal solution for $Z$ in terms of $X,Y$, in general, is provided by the {\it Baker-Campbell-Hausdorff formula} (BCH):   
\begin{equation} \label{BCH} 
Z=X+Y+\frac{1}{2}[X,Y]+\frac{1}{12}
([X,[X,Y]]+[Y,[Y,X]])  
-\frac{1}{24}[Y,[X,[X,Y]]]
+\;\dots
\;\;. \end{equation}  
The coefficients in this series of increasingly complicated iterated commutators are known.  However, to assess the convergence or divergence of the series, generally, turns out to be 
difficult. There are exceptional cases, when the series terminates and in recent works by Visser {\it et al.} and by Matone -- see also the numerous references there -- new types of such closed form solutions have been studied \cite{Visser,Matone}.    

Returning to Eq.\,(\ref{U}), we notice that individual pair exchange permutations composing the evolution operator $\hat U$ can be exponentiated easily: 
\begin{equation} \label{exp} 
\hat P_{ij}=i\exp (-i\frac{\pi}{2}\hat P_{ij}) 
\;\;, \end{equation} 
since $(\hat P_{ij})^2=\mathbf{1}$, {\it cf.} Eqs.\,(\ref{Pijdef}). However, 
because of their noncommutativity, Eq.\,(\ref{comm}), it was {\it not} possible to calculate the Hamiltonian $\hat H$ by simply adding all resulting exponents.  

Using Eq.\,(\ref{U}) (with the respective factors {\it within} first and second product  commuting), Eq.\,(\ref{H3}), and Eq.\,(\ref{exp}), we arrive here at the following equalities:  
\begin{eqnarray}\label{BCH1} 
\hat U&:=&\prod_{k=1}^{S}\hat P_{2k\;2k+1}\prod_{l=1}^S\hat P_{2l-1\;2l}
\nonumber \\ [1ex] \label{BCH2}
&=&i^{2S}\exp \big (-i\frac{\pi}{2}\sum_{k=1}^{S}\hat P_{2k\;2k+1}\big )
\exp \big (-i\frac{\pi}{2}\sum_{l=1}^{S}\hat P_{2l\;2l+1}\big ) 
\\ [1ex] \label{BCH3} 
&=&\exp \Big (-i\pi  
\big (\mathbf{1}
+\frac{i}{S}\sum_{n=1}^{S-1}\cot (\frac{\pi}{S}n )\hat U^{n}\big )\Big )
\;\;,\;\mbox{for}\;\;\hat U|\psi\rangle\neq |\psi\rangle 
\;\;. \end{eqnarray} 
For the zero modes, with $\hat U|\psi_0\rangle =|\psi_0\rangle$, a trivial identity results. 
Instead, between the right-hand sides of Eqs.\,(\ref{BCH2}) and (\ref{BCH3}), we find a 
rather involved yet {\it terminating BCH formula}.\footnote{Overall, beginning with $\hat U$ itself on the first line, the relations produce also an interesting functional equation for $\hat U$.} Note that the final expression on the right-hand side of Eq.\,(\ref{BCH3}) can be factorized into exponentials, since only commuting powers of $\hat U$ are involved. 

\section{``Ontological Ising spins + $\epsilon$ = quantum spin chain''} 

We have noticed earlier in Refs.\,\cite{ElzeQu19,ElzePAFT19} that the 3- and 4-spin chains,  considered as most simple ontological models -- {\it i.e.} classical and deterministic -- tend to appear as of genuinely quantum mechanical kind. We shall illustrate this 
surprising feature in more detail here. 

To begin with, the Baker-Campbell-Hausdorff formula, Eqs.\,(\ref{BCH1})--(\ref{BCH3}), which relates the ontological updating operator $\hat U$ to the corresponding Hamiltonian $\hat H$, 
{\it cf.} Eqs.\,(\ref{H1})--(\ref{H4}), is characterized by precisely determined numerical 
coefficients, which figure here as sort of fine tuned {\it coupling constants}.\footnote{It follows from Eq.\,(\ref{exp}) that the coefficients $\pi /2$ can be replaced on the right-hand side of Eq.\,(\ref{BCH1}) without harm by $(2k+1/2)\pi$, with integer $k$. 
Similarly, a substitution by $(2k+3/2)\pi$ can be absorbed.}  

Next, we may think of classical Ising spins as embedded in the larger Hilbert spaces of corresponding quantum spins (or qubits). Then, the exchange operators $\hat P_{ij}$, Eq.\,(\ref{PijPauli}), followed 
by the unitary $\hat U$, Eq.\,(\ref{U}), and finally the Hamiltonian are all expressed in terms of Pauli matrices as proper {\it quantum mechanical operators}. Yet, these describe exactly the {\it classical dynamics} of the ontological Ising spins generated by exchange interactions. However, this all holds, {\it if and only if} the numerical constants in $\hat H$ have 
the precise values from Eqs.\,(\ref{H1})--(\ref{H4}). -- It is worth noting that the Hamiltonian contains highly nonlocal terms, {\it e.g.} $\sim\hat U^{S-1}$, which address all spins in the chain. Nevertheless, these terms are small, {\it cf. below}.  

In realistic situations, with physical theories and their parameters induced by experimental findings, the {\it coupling constants}, such as the parameters of the Standard Model, are not precisely known.\footnote{Which is one motivation for the search of physics beyond the Standard Model, namely to yield stronger constraints on the large set of its parameters.} 

Therefore, it is plausible that the apparent {\it ``quantum instability''} of the underlying ontological model arises, when the {\it coupling constants} are imprecise or small terms are missing in the 
deduced Hamiltonian. In this situation we are forced to use the full quantum mechanical apparatus in the description of what is observed, instead of the still unknown ontological model. 

In the present example of the classical Ising spin chain {\it small perturbations} of the fixed numerical constants turn the Hamiltonian into an operator that likely produces {\it superpositions of multi-spin states}, a hallmark of quantum mechanics. The exact Hamiltonian instead, as above, produces permutations of {\it ontological states}, by construction.   

Using the asymptotic behaviour of the $\cot$-function, for a long spin chain with $S\gg 1$, and 
in particular,  
$$\cot (\frac{\pi}{S}n)=\pm\frac{S}{\pi}\Big (1+\mbox{O}\big ((\pi /S)^2\big )\Big )\;\;,\; 
\mbox{for}\;\; n=\Big\{
\begin{array}{c}
1 \\ S-1 \end{array} \;\;,$$ 
we approximate the Hamiltonian by the leading terms. Thus, from Eq.\,(\ref{H4}): 
\begin{equation}\label{Happrox}
\hat H\approx \frac{\pi}{T}\Big (\mathbf{1}
+\frac{i}{2\pi}
\big (\hat U-\hat U^\dagger 
-\hat U^{S-1}+(\hat U^\dagger )^{S-1}
\big )\Big )
=\frac{\pi}{T}\Big (\mathbf{1}
+\frac{i}{\pi}
\big (\hat U-\hat U^\dagger 
\big )\Big )
\;\;, \end{equation} 
using $\hat U\hat U^\dagger =\mathbf{1}=\hat U^S$. In line with the previous discussion, this could arise in a fictitious world, where only these leading terms are inferred from first observations of an ontological Ising spin chain, such as studied here. 

Obviously, the approximate Hamiltonian of Eq.\,(\ref{Happrox}) produces {\it superpositions}, when 
applied to ontological states, such as $|\psi\rangle$ of Eq.\,(\ref{psi}), which are not ontological states any longer. In particular, it is easy to see that $\hat U^\dagger$ acts like $\hat U$ on such a state, however, with all  
directions of motion reversed, {\it cf.} the discussion of Eqs.\,(\ref{psi})--(\ref{Uspsi}). 

Let us consider an example, where $\hat H$, approximated as above in Eq.\,(\ref{Happrox}), is applied to a particular ontological state. We choose: 
\begin{equation}\label{upup} 
|\psi_{\downarrow\downarrow}\rangle :=|\;\dots\;,\uparrow ,\uparrow ,\uparrow ,\downarrow ,
\downarrow ,\uparrow ,\uparrow ,\uparrow ,\;\dots\;\rangle 
\;\;, \end{equation} 
assuming that the first encountered {\it down}-spin from the left is on an {\it even} numbered site, for definiteness; all spins not explicitly shown are assumed to be in the {\it up}-state. Then, we obtain, using the previous results: 
\begin{eqnarray} 
\big (\hat U-\hat U^\dagger \big )|\psi_{\downarrow\downarrow}\rangle 
&=&\;\;\;|\;\dots\;,\uparrow ,\;\uparrow ,\downarrow ,\;\uparrow , \uparrow ,\;\downarrow ,\uparrow ,\;\uparrow  ,\;\dots\;\rangle  
\nonumber \\ [1ex] \label{ent} 
&\;&-|\;\dots\;,\uparrow ,\;\downarrow ,\uparrow ,\;\uparrow , \uparrow ,\;\uparrow ,\downarrow ,\;\uparrow  ,\;\dots\;\rangle  
\;\;. \end{eqnarray} 
The interesting point here is that this part of the Hamiltonian {\it locally} creates structures that are reminiscent of an {\it entangled Bell state} of two quantum spins, such as  $\sim |\uparrow\downarrow -\downarrow\uparrow\rangle$. -- This raises some interesting questions. Can the Bell state structures, visible in Eq.\,(\ref{ent}), move far from each other, possibly depending on interactions, giving rather isolated ``local Bell states''? 
What is a quantitative measure of the entanglement in the state of Eq.\,(\ref{ent})? We have to  come back to this eventually.   

To summarize, the classical deterministic Ising spin-chain that we have studied in this work has proven to be a successful testing ground for ideas concerning ontological models underlying quantum mechanics \cite{tHooft2014,tHooft2020}. The ontological dynamics-from-permutations has been shown here to explain the need for the quantum mechanical description, when ontological degrees of freedom and deterministic laws governing their motion are not perfectly known.     

\section{Conclusions} 

We presently extend recent considerations which attempt to illustrate the {\it Cellular Automaton Interpretation of Quantum Mechanics} \cite{tHooft2014}, by generalizing from three- and 
four-spin ontological models \cite{ElzeQu19,ElzePAFT19} to an arbitrarily long chain of $2S$ interacting classical Ising spins. They evolve deterministically, controlled by specifically chosen exchange permutations among the $2^{2S}$ states. 

In the Introduction we provide a synopsis of this new interpretation of quantum mechanics, which should provide a shortcut to essentials for an uninitiated reader.\footnote{\dots who might suffer with us from the all-too-often blurred terminology when it comes to the foundations of quantum theory.} 

We recall results obtained for so-called cogwheel models in Section\,2, which describe systems that periodically evolve through a discrete set (of possibly a large number) of $N$ states \cite{tHooft2014,tHooft2020,Elze,ElzeRelativPart}. 
This is related to properties of a unitary $N\times N$ {\it permutation matrix in standard form}. Considering this as update operator for the evolution {\it via} a finite time step, we  
obtain the corresponding Hamiltonian operator, its eigenvalues, and its eigenstates. 

This kind of model, so far, did not require any information on the nature of the ontological states $\cal OS$ nor on their dynamics, besides resulting from complete permutations (excluding dissipative modifications for now). Therefore, it may have a wide range of applications. 

In Section\,3, we introduce permutations realized by Ising spin (or {\it bit}) exchanges. Then, we apply results of Section\,2 to extract the corresponding Hamiltonian, resulting in Eq.\,(\ref{H4}). It is characterized by a highly degenerate spectrum, which is related to conservation laws that cause the evolution operator to have a block diagonal structure, and a high degree of nonlocality. Nevertheless, the model has a {\it finite signal velocity} (``velocity of light'') and a first interpretation, in Section\,3.3.2, brings it close to a {\it free field theory}. 

It will be interesting to understand large-scale limits of the model, either 
by applying {\it Sampling Theory} (straightforward), as for example in Ref.\,\cite{PRA2014}, or mean field theory, or some decimation and coarse graining procedure (difficult, in view of the structure of the Hamiltonian).  

We showed that our results for the Hamiltonian can be concisely rewritten to generate a new {\it Baker-Campbell-Hausdorff formula} with terminating expansion in terms of permutation operators, {\it cf.} Eqs.\,(\ref{BCH})-(\ref{BCH3}). 

The Ising spin chains treated here are intrinsically classical deterministic systems. Yet their description using appropriate quantum mechanical language offers several similarities with genuine quantum systems. Most importantly, we have observed their {\it ``quantum instability''}, which is visible in the Baker-Campbell-Hausdorff formula or in the Hamiltonian itself. Small perturbations can spoil these results and will, generally, lead to dynamics that produces superpositions of {\it quantum spins} or {\it qubits}. Which are no longer $\cal OS$. In this way, underlying classical ontological features are replaced by what appears ``naturally'' to be described by quantum theory. 

{\it Here ontological 
Ising spin systems appear as islands in parameter space embedded in a large sea of quantum spin models}.   
 
The crucial next step, in bringing the presented ontological models closer to what quantum field theory in all its glory offers, will consist in understanding how such spin chains fuse or break, possibly incorporating a mechanism analogous to the interaction of different cogwheels outlined in Ref\,\cite{tHooft2020}. 

\begin{acknowledgements}
It is a great pleasure to thank Andrei Khrennikov for instigating the writing of this paper. 
\end{acknowledgements}

%\section*{Conflict of interest}
%The authors declare that they have no conflict of interest.

\end{document}